\documentclass[prb,aps,twocolumn,showpacs]{revtex4}
\usepackage{graphicx}
\begin{document}

\title{Mechanisms of Manganese-Assisted Nonradiative Recombination in Cd(Mn)Se/Zn(Mn)Se Quantum Dots}

\author{A.~V.~Chernenko}
\address{Institute of Solid State Physics,RAS,
142432,Chernogolovka, Russia}
\email{chernen@issp.ac.ru}
\author{A.~S.~Brichkin}
\address{Institute of Solid State Physics,RAS,
142432,Chernogolovka, Russia}
\author{N.~A.~Sobolev}
\address{Departamento de Fi'sica and I3N, Universidade de
Aveiro,3810-193 Aveiro, Portugal}
\author{M.~C.~Carmo}
\address{Departamento de Fi'sica and I3N, Universidade de
Aveiro,3810-193 Aveiro, Portugal}

\begin{abstract}

Mechanisms of nonradiative recombination of electron-hole complexes in Cd(Mn)Se/Zn(Mn)Se quantum dots accompanied by interconfigurational excitations of Mn$^{2+}$ ions are analyzed within the framework of single electron model of deep {\it 3d}-levels in semiconductors. In addition to the mechanisms caused by Coulomb and exchange interactions, which are related because of the Pauli principle, another mechanism due to {\it sp-d} mixing is considered. It is shown that the Coulomb mechanism reduces to long-range dipole-dipole energy transfer from photoexcited quantum dots to Mn$^{2+}$ ions. The recombination due to the Coulomb mechanism is allowed for any states of Mn$^{2+}$ ions and {\it e-h} complexes. In contrast, short-range exchange and ${\it sp-d}$ recombinations are subject to spin selection rules, which are the result of strong {\it lh-hh} splitting of hole states in quantum dots. Estimates show that efficiency of the {\it sp-d} mechanism can considerably exceed that of the Coulomb mechanism. The phonon-assisted recombination  and processes involving upper excited states of Mn$^{2+}$ ions are studied. The increase in PL intensity of an ensemble of quantum dots in a magnetic field perpendicular to the sample growth plane observed earlier is analyzed as a possible manifestation of the spin-dependent recombination.
\end{abstract}

\maketitle

\section{Introduction}

A great deal of attention attracted to quantum dot (QD) structures is due to the possible use of their quantum states in various fields of spintronics and for generation and detection of light.\cite{Awschalom} Semimagnetic (diluted magnetic semiconductor (DMS)) II-VI quantum dots are promising objects for these purposes because a high degree of spin polarization of electrons and holes can be achieved in relatively weak magnetic fields.

However, incorporation of Mn ions into CdSe/ZnSe QDs substantially reduces the quantum yield of radiation as soon as the optical transition energy exceeds the energy of internal Mn transition $\sim 2.15$ eV. It is found that Cd(Mn)Se/Zn(Mn)Se QD photoluminescence(PL) is completely quenched due to nonradiative recombination at relatively small Mn content $x \simeq 3-5\%$. \cite{PRB05,Henneberger01,Kim98,Makowski02,Lee05}

The PL quenching is due to nonradiative recombination of QD {\it e-h} complexes accompanied by excitation of  Mn$^{2+}$ ions. The Coulomb interaction between {\it 3d-} and band electrons is usually considered as a reason for processes of energy transfer from {\it e-h} complexes to Mn$^{2+}$ ions. \cite{Robbins,Allen86}

The interest in possible mechanisms of energy transfer arose after observation of a strong increase in PL intensity of an ensemble of Cd(Mn)Se/Zn(Mn)Se QDs in magnetic field ${\bf B}\parallel {\bf 0z}$ reported by many authors. \cite{PRB05,Henneberger01,Kim98,Makowski02,Lee05} Simultaneous increase in the QD  PL life-time was also reported. \cite{Makowski02,Falk02,Falk03}

The explanation of the increase is based on suppression of the spin-dependent nonradiative recombination of {\it e-h} pairs by magnetic field. According to the model proposed in Ref.\cite{Nawrocki95} nonradiative exciton recombination is possible because of the direct exchange interaction between band carriers and {\it 3d} electrons. This model provides selection rules for the process: $S'_{B}=S_{B}$, where $S_{B}$ and $S'_{B}$ are Mn$^{2+}$ spin projections in the direction of  magnetic field $B$ in the initial and final Mn$^{2+}$ states respectively. Selection rules predict that the nonradiative recombination of bright excitons is forbidden for the states of Mn$^{2+}$ ions with  $S_{B}=\pm 5/2$ whereas it is allowed for Mn$^{2+}$ ions with other $S_{B}$. Since the number of Mn$^{2+}$ ions with $S_{B}=-5/2$ increases with magnetic field, nonradiative recombination is suppressed.  The rules correctly explained the strong increase in PL intensity in Cd(Mn)Se/Zn(Mn)Se QDs but failed to explain its dependence of the direction of the magnetic field. \cite{PRB05} To explain the observed dependence a modification of selection rules was suggested in Ref.\cite{PRB05}  $S_{z}+s_{ex,z}=S'_z$, where $S_{z}$ is the projection of Mn$^{2+}$ spin in {\bf 0z} axis, which is perpendicular to the sample growth plane, $s_{ex,z}=s^e_z+s^h_z$ is the projection of exciton spin (instead of the total angular momentum $J_z=j^h_z+s^e_z$) in {\bf 0z}. This suggestion is based on PL studies of Cd(Mn)Se/Zn(Mn)Se and CdSe/ZnSe/ZnMnSe QDs, which revealed that {\it hh} and {\it lh} hole states are strongly split due to strain and dimensional quantization. \cite{PRB07,JETP07} The carriers in upper {\it hh} band are characterized by both the moment projection $j^h_z=\pm 3/2$ and the spin projection $s^h_z=\pm 1/2$. The Coulomb and exchange mechanisms are assumed to obey the same selection rules. \cite{PRB05}

The analysis presented below shows, however, that spins of atomic configurations were not properly regarded in Ref.\cite{PRB05} and some correction of previous results is required.  Besides, the {\it sp-d} mixing specific to DMS was omitted while its contribution to the nonradiative recombination can be essential. The consideration based on the single-electron model of the {\it 3d}-level in a semiconductor reveals that mechanisms based on the direct exchange interaction and  {\it sp-d} mixing are subject to the spin selection rules  In contrast, the Coulomb mechanism does not lead to the spin-dependent recombination.   Estimates show that  efficiency of the {\it sp-d}  mechanism can noticeably exceed efficiency of the Coulomb process in Cd(Mn)Se/Zn(Mn)Se QDs. The {\it sp-d} mechanism can either contribute to impact excitation of  Mn$^{+2}$ ions incorporated into a II-VI semiconductor matrix, which is closely related to the nonradiative recombination.

The paper is organized as follows. In the next section a model Hamiltonian of a DMS QD structure is considered. Mechanisms of the nonradiative recombination due to the Coulomb interaction and {\it sp-d} mixing are analyzed in Sec. III. Phonon-assisted processes and energy transfer into upper Mn$^{2+}$ excited states are discussed in Sec.IV. The link between the model in question and the impact excitation of Mn$^{2+}$ ions is also discussed in Sec.IV. Spin-dependent selection rules for nonradiative recombination and their manifestations in experiments are considered in Sec.V.

\section{Model Hamiltonian}

The Hamiltonian of the structure containing one QD and one Mn$^{2+}$  ion has the form of an Anderson Hamiltonian: \cite{Larson88,Hass,Young91}
\begin{equation}
\hat H_0=\hat H_{QD}+\hat H_{Mn}+\hat H_{hyb}+\hat H_x,
\label{Hamiltonian}
\end{equation}
where $\hat H_{QD}$ is the Hamiltonian of the QD electron system, $\hat H_{hyb}$ is the hybridization Hamiltonian, $\hat H_{Mn}$ is the Hamiltonian of the Mn$^{2+}$ {\it 3d}-shell  and the Hamiltonian of the potential exchange interaction is $\hat H_x$.

The QD Hamiltonian
\begin{equation}
\hat H_{QD}=\hat H_c+ \hat H_v+\hat H_{cv}
\end{equation}
contains Hamiltonians of the conduction  $\hat H_c=\sum_{\mu s_z} \epsilon_{\mu s_z}c^+_{\mu s_z}c_{\mu s_z}$ and valence $\hat H_v=\sum_{\nu j_z} \epsilon_{\nu j_z}b^+_{\nu j_z}b_{\nu j_z}$ electrons. Here $\mu$ and $\nu$ enumerate electron and hole states in QD. Conduction and valence electron states are $|\varphi^e_{\mu s_z}\rangle=c^+_{\mu s_z}|0\rangle$ and $|\varphi^v_{\nu j_z}\rangle=b^+_{\nu j_z}|0\rangle$, respectively.  Electron, hole, exciton, and trion states are eigen-states of $\hat H_{QD}$. Hole states are related to valence electron states via a time-reversal operation $|\varphi^h_{\nu j_z}\rangle =|\varphi^{v*}_{\nu -j_z}\rangle$.

The term $H_{cv}$ contains {\it e-h} Coulomb and exchange interactions. The Coulomb term mixes single-electron states and leads to the reduction of the {\it e-h} energy whereas the {\it e-h} exchange term splits {\it e-h} states with $J=1,2$. Without loss of generality we retain only the exchange term and neglect the Coulomb interaction  because  the confinement of {\it e-h} states in QDs is determined by the dimensional quantization and because  its contribution to energy transfer is negligible as it will be clear in Sec.III.

The dots under study have a very anisotropic lens-like form, i.e. the diameter $D$ is several times larger than the height $L$. Photoluminescence studies of Cd(Mn)Se/Zn(Mn)Se and CdSe/ZnSe/ZnMnSe QDs  reveal that most of QDs have symmetry $C_{2v}$. These studies show that {\it hh} and {\it lh} hole states are strongly split due to strain and dimensional quantization. \cite{PRB07,JETP07} The {\it e-h} exchange interaction splits {\it e-h} states with $J_z=\pm 1$ ("bright excitons") and $J_z=\pm 2$ ("dark excitons"). The gap between bright and dark states is around 2-3 meV. \cite{PRB07}  The wave-functions of bright exciton in those structures are $|\psi^b_{ex}\rangle =1/\sqrt{2}(|1\rangle\pm |-1\rangle)$, where $|\pm 1\rangle$ are exciton states with $J_z=\pm 1$ and $|\psi^d_{ex}\rangle=|\pm 2\rangle$ are dark exciton  states.

The effect of the periodic lattice potential on band electrons is taken into account within the effective-mass approximation, which may be used since we are interested in the properties of states near the bottom of the conduction and top of the valence bands. In the effective mass approximation conduction electron states are  $\varphi^e_{s_z}({\bf r})=F_{es~s_z}({\bf r})S_{s_z}+ F_{ex~s_z}({\bf r})X_{s_z}+F_{ey~s_z}({\bf r}) Y_{s_z}+F_{ez~s_z}({\bf r})Z_{s_z}$, where envelopes  $F_{es~s_z}({\bf r}),F_{ex~s_z}({\bf r}),F_{ey~s_z}({\bf r}),F_{ez~s_z}({\bf r})$ are solutions of the effective mass equations. Here $S$, $X$, $Y$, $Z$ are zone center Bloch functions of appropriate symmetries. The electron states on the bottom of the first band of dimensional quantization within symmetric quantum well of width $L_w$ can be found elsewhere. \cite{Merkulov00}  Bloch amplitudes of free electrons with ${\bf k}\parallel {\bf 0z}$ that account for $k^2$ terms within  the Kane's model  are $\varphi^e_{\pm 1/2}(k)=(1-(\hbar kp)^2/6m^2_0)S_{\pm 1/2}\pm p\hbar k/(3m_0)[(1/E_g-1/(E_g+\Delta))(X\pm iY)_{\mp 1/2}+(2/E_g+1/(E_g+\Delta))Z_{\pm 1/2}]$. \cite{Merkulov99} Here  $p=\langle S|p_x|X\rangle=iPm_0/\hbar$, $P$ is the Kane parameter, $m_0$ is the free electron mass. \cite{Abakumov} These results show that the admixing of Z functions to $\varphi^e_{s_z}({k})$ is much larger than those of X and Y functions.

The admixture of $lh$ to $hh$ states takes place in QDs of $C_{2v}$ symmetry and lower.\cite{JETP07} However, it is small and unimportant for our aims. Therefore, we assume that $\varphi^h_{\nu \pm 3/2}({\bf r})=\varphi^{v*}_{\nu \mp 3/2}({\bf r})= \pm 1/\sqrt{2}F_{hh\nu}({\bf r})(X\pm iY)_{\pm 1/2}$, where basis functions from Ref.\cite{Abakumov} are used. Thus, QD exciton states are $|\pm 1\rangle=|\varphi^h_{\nu \pm 3/2}\rangle|\varphi^e_{\mp 1/2}\rangle$ and $|\pm 2\rangle=|\varphi^h_{\nu \pm 3/2}\rangle|\varphi^e_{\mu \pm 1/2}\rangle$.

The Hamiltonian of the Mn$^{2+}$ ion in the Hubbard form is
\begin{equation}
\hat H_{Mn}=\sum_{ms_z}\hat n_{ms_z}(\epsilon_{d}+U_{eff}\hat n_{m-s_z}),
\end{equation}
where $\hat n_{ms_z}$ is the number operator of {\it 3d}-electrons with the orbital momentum $m$ and spin projection $s_z$, $U_{eff}$ is the electrostatic repulsion energy between {\it 3d}-electrons occupying the same state $m$. We assume that $\hat H_{Mn}$ acts only on $d^4$, $d^5$, $d^6$ configurations of Mn$^{2+}$ ions containing n=4,5, and 6 {\it 3d}-electrons, respectively. \cite{Larson88,Schriffer67} The crystal field splitting of {\it 3d}-shell is omitted in $\hat H_{Mn}$. When it is taken into account the Hamiltonian can be expressed as $\hat H_{Mn}=\sum_i|A_i\rangle \epsilon_i\langle A_i|$, where $|A_i\rangle$ and $\epsilon_i$, respectively, are states and energies of $d^4$-$d^6$ configurations. They can be found by means of the crystal field theory. \cite{Griffith}

The hybridization Hamiltonian
\begin{equation}
\hat H_{hyb}=\hat H_{sd}+\hat H_{pd}
\end{equation}
consists of $\hat H_{pd}=\sum_{ms_z,\nu j_z}(V_{pd~ms_z,\nu j_z}d^+_{ms_z}b_{\nu j_z}+h.c.)$ and $\hat H_{sd}=\sum_{m s_z,\mu s'_z}(V_{sd~ms_z,\mu s'_z}d^+_{ms_z}c_{\mu s'_z}+h.c.)$ terms. The {\it sp-d} hybridization results from the combined influence of the potential of crystal ions and band  electrons. \cite{Griffith} It is considered on the symmetry grounds as a result of action of an effective crystal field potential $\hat U_{cr}$, so that hybridization coefficients $V_{sd~ms_z \mu s'_z}=\langle d_{ms_z}|\hat U_{cr}|\mu s'_z\rangle$ and $V_{pd~m s_z\nu j_z}=\langle d_{ms_z}| \hat U_{cr}|\nu j_z\rangle$. In the case of substitutional Mn$^{2+}$ ions the crystal field mixes valence band states at ${\bf k}\simeq 0$ only with {\it 3d}-functions, which belong to $t_2$ representation of the $T_d$ site symmetry. \cite{Bhatt92,Blinowski91,Kikoin} The operator $\hat U_{cr}$ also mixes {\it 3d}- and conduction band states at ${\bf k}\neq 0$. \cite{Merkulov00}

The direct {\it s-d} exchange term is
\begin{equation}
\hat H_x=-{\sum_\lambda}J({\bf R}_\lambda){\bf S}_\lambda\sum_{\nu s_z,\nu's'_z}  c^+_{\nu s_z} {\bf\hat s}_{ s s'_z}c_{\nu' s'_z},
\label{exchange}
\end{equation}
where $J({\bf R}_\lambda)$ is the exchange constant, ${\bf S}_\lambda$ is the spin of the $\lambda$-th Mn$^{2+}$ ion located at ${\bf R}_\lambda$, ${\bf \hat s}$ is the electron spin operator. The direct exchange is associated with the Coulomb interaction between band and {\it 3d}-electrons. The direct {\it p-d} exchange is considered to be zero. \cite{Larson88,Hass,Merkulov99}

Reduction of the PL quantum efficiency in semiconductor structures due to presence of impurities is usually related to the nonradiative recombination of {\it e-h} pairs because of the Coulomb or exchange interactions between band and impurity states. \cite{Allen86,Abakumov,Yassievich93}  The Coulomb interaction is missed in the Hamiltonian in Eq.(\ref{Hamiltonian}).  In order to account for it the approach developed in Ref.\cite{Kikoin} is used.

The Anderson Hamiltonian in Eq.(\ref{Hamiltonian}), which was introduced phenomenologically into the DMS theory, \cite{Larson88,Hass} can be obtained as a generalization of the Slater-Koster equation for single-electron states in a crystal containing an impurity level $(H_0+H_{imp}-E)\phi_e({\bf r})=0$, where $H_0$ is the single-electron Hamiltonian in an ideal crystal structure in the absence of the impurity, and $H_{imp}$ is the potential of impurity. \cite{Kikoin} The single-electron states are expressed via eigen-states $\phi_i({\bf r})$ of $H_0$  $\varphi_e({\bf r})=\sum_i F_i\phi_i({\bf r})$, where $F_i$ are appropriate coefficients and summation is made over quantum numbers characterizing crystal states.

The single-electron potential of perfect crystal structure may be written as   $V^0=\sum_\lambda v_\lambda^0({\bf r}-{\bf R}^0_\lambda)+V^0_e({\bf r})$, \cite{Pantelides} where $V^0_e({\bf r})$ is the potential due to band electrons, $v_\lambda^0({\bf r} - {\bf R}^0_\lambda)$ is the potential of core electrons and nucleus at ${\bf R}^0_\lambda$. In the presence of impurity the potential may be written as $V=\sum_\lambda v_\lambda({\bf r}-{\bf R}_\lambda)+V_e({\bf r})$, where ${\bf R}_\lambda$ are new atomic sites. Therefore the perturbation is\\

$H_{imp}=V-V^0=\sum_\lambda (v_\lambda({\bf r}-{\bf R}_\lambda)-v_\lambda^0({\bf r}-{\bf R}^0_\lambda))+U_c({\bf r})\simeq
$
\begin{equation}
\sum_\lambda (v_\lambda({\bf r}-{\bf R}^0_\lambda)-v_\lambda^0({\bf r}-{\bf R}^0_\lambda))-
\sum_\lambda\left(\frac{\partial v_\lambda({\bf r}-{\bf R}_\lambda)}{\partial {\bf R}_{\lambda}}\right)_0
\end{equation}
$$\times({\bf R}-{\bf R}^0_\lambda)+U_c({\bf r}),$$
where $U_c({\bf r})=V_e({\bf r})-V^0_e({\bf r})$.

By neglecting lattice relaxation (${\bf R}_\lambda={\bf R}^0_\lambda$) the potential of a substitutional impurity at ${\bf a}_0$ may be written as $H_{imp}({\bf r})=v_{3d}({\bf r}-{\bf a}_0)-v_h({\bf r}-{\bf a}_0)+U_c({\bf r})$, \cite{Kikoin}  whereas the potential of an interstitial impurity is $H'_{imp}({\bf r}-{\bf a}_0)=v_{3d}({\bf r}-{\bf a}_0)+U_c({\bf r})$. Here $v_h({\bf r}-{\bf a}_0)$ is the potential of the host ion. Only substitutional impurities are discussed below since obtained results can be easily extended to the latter case.

The Hamiltonian of the crystal structure with substitutional impurity in the second-quantized form can be expressed as $\hat H=\hat H_1+\hat H_2$, where  $\hat H_1=\hat H_0+\hat H_{imp}$  is the single-electron Hamiltonian in the basis $\{d_{ms_z},\phi^e_{\nu s_z},\phi^v_{\mu j_z}\}$, that includes localized {\it 3d}-states $d_{ms_z}$. The term $\hat H_2$ contains interactions between single-electron states. The only interaction included in the Anderson model is the Coulomb repulsion between electrons with the same $m$ but opposite $s_z$ $\hat H_2=U_{eff}\sum_{ms_z}\hat n_{ms_z}\hat n_{m-s_z}$. Besides  $\hat H_{QD}$, $\hat H_{hyb}$ and $\hat H_{d}=\sum_{ms_z}\epsilon_d d^+_{ms_z}d^+_{ms_z}$ the Hamiltonian $\hat H_1$  contains the scattering term

\begin{equation}
\hat H_{scat}=\sum_{\mu \neq \mu',\nu \neq \nu', s_z,j_z}(U_{\nu s_z,\nu' s_z}c^+_{\nu s_z}c_{\nu' s_z}+
U_{\mu j_z,\mu' j_z} b^+_{\mu j_z}b_{\mu' j_z})
\end{equation}
$$+\sum_{\mu,\nu, s_z,j_z}(U_{\nu s_z,\mu j_z} c^+_{\nu s_z}b_{\mu j_z}+h.c.)
$$
usually omitted in the canonical Anderson model. Here $U_{\nu s_z(\mu j_z),\nu' s'_z(\mu' j_z)}=\langle \nu s_z(\mu j_z)|H_{imp}|\nu' s'_z(\mu' j_z) \rangle$.

The Bloch wave-functions  $\Psi_{vs_z}e^{i{\bf kr}}$, where $\Psi_{vs_z}$ can be either  $X_{s_z}$, $Y_{s_z}$, $Z_{s_z}$ or $S_{s_z}$, are not orthogonal to $d_{ms_z}$ at ${\bf k} \neq 0$.  As a result anticommutators  $[d_{\mu s_z}c^+_{\mu s_z}]=\langle d_{ms_z}|\varphi^{e}_{\mu s_z}\rangle \neq 0$ and $[d_{m s_z}b^+_{\nu j_z}]=\langle d_{ms_z}|\varphi^{v}_{\nu j_z}\rangle \neq 0$ because slowly varying envelopes $F_{es,ex,ey,ez,hh}({\bf r})$ can always be expressed as  $\sum_{\bf k} A({\bf k})e^{i{\bf kr}}$. We assume, however, that overlap integrals are negligible because we are interested in band states near ${\bf k}=0$.

To account for the {\it sp-d} Coulomb interaction between band and five {\it 3d}-electrons the terms  originating from  $\sum^5_{i=1} e^2/|{\bf r}-{\bf r}_i|$ should be added to $\hat H_2$. The potential $U_c({\bf r})$ tends to compensate changes in electron density caused by  presence of the impurity and leads to the screening of the interaction between band and 3d-electrons. Within the framework of linear response theory  $H_{imp}({\bf q})= 1/\epsilon_{eff}({\bf q})(u_d({\bf q})-u_h({\bf q}))$, where $H_{imp}(\bf q)$, $u_d(\bf q)$, $u_h(\bf q)$  are Fourier transforms of $H_{imp}({\bf r})$, $u_d(\bf r-{\bf a}_0)$, $u_h({\bf r}-{\bf a}_0)$, respectively. \cite{Pantelides}
Similarly, the screened Coulomb interaction between band and {\it i}-th 3d-electron is

\begin{equation}
v_{sc}({\bf r}-{\bf r}_i)=\frac{4\pi e^2}{(2\pi)^3}\int \frac{e^{i{\bf q(r-r_i)}}}{\epsilon_{eff}(\bf q){\bf{q}^2}} d^3{q}
\label{Coul}
\end{equation}

This expression is appropriate for the Coulomb processes of nonradiative recombination that are characterized by small  transferred quasi-momentum $q_{\parallel}\simeq 1/D$, $q_z\simeq 1/L$.   Here $D$ and $L$ are characteristic lengths of QDs in the lateral and {\bf 0z}-direction, respectively. Quasi-momentums within this range participate in the formation of {\it e-h} states, which justifies the use of the linear response approach. \cite{Pantelides} Since  the transferred energy $E_0\simeq $2.15 eV satisfies condition $E_0\simeq E_g \gg\hbar\omega_{LO}$, the inert ionic system of the crystal does not contribute to the dielectric function so that the effective dielectric function is determined by band electrons. The processes caused by the long-range Coulomb interaction are characterized by $q\simeq 0$  so that for the Coulomb processes $\epsilon_{eff}({\bf q\simeq 0})\simeq \epsilon_{\infty}$.

Contrary to the Coulomb processes exchange ones are characterized by a small interaction radius so that contribution of the processes with large transferred quasimomentums can be noticeable. The explicit form of the screened Coulomb interaction is not required for the future analysis of the exchange mechanism so that the expression $v_{sc}({\bf r}_1-{\bf r}_2)$ is used in this case.

The Coulomb interaction between band and {\it 3d}-electrons $V_{sc}=\sum^5_{i=1} v_{sc}({\bf r}-{\bf r}_i)$
generates many terms of the form $V_{i,j,l,k}a_i^+a_j^+a_ka_l,$ where $a^+_i(a_i)$ can be either of $d^+_{ms_z}(d_{ms_z}),c^+_{\mu s_z}(c_{\mu s_z}), b^+_{\nu j_z}(b_{\nu j_z})$. The terms containing only {\it 3d-} or band electron operators, i.e. those that do not couple {\it 3d-} and band states, are not considered here since the effect of these terms are already accounted for in $\hat H_{Mn}$ and $\hat H_{QD}$.

The other terms  appeared because of $H_{imp}+V_{sc}$ are analyzed by means of the approach developed in Ref. \cite{Elliot86}   The analysis shows that some terms originated from $V_{sc}$ are already included in $\hat H_0$.

At distances $|{\bf r}-{\bf a}_0| \gg a_B$, where $a_B$ is the atomic Bohr radius,
$v_{sc}({\bf r}-{\bf r}_1)\simeq v_{sc}({\bf r}-{\bf a}_0)$, so that
the term
\begin{equation}
\sum_{ms_z,p,q} V_{p,ms_z,q,ms_z}\hat n_{m s_z}a^+_pa_q
\end{equation}
tends to compensate $\sum_{p,q} U_{p,q} a^+_pa_q=\sum_{p,q}\langle a_p|H_{imp}({\bf r}-{\bf a}_0)|a_q\rangle a^+_pa_q$ that can be expressed as  $\sum_{p,q}\langle a_p|Z_{Mn}e^2/\epsilon_{eff}|{\bf r}-{\bf a}_0||a_q\rangle a^+_pa_q$, where $Z_{Mn}$ is the net core charge that includes Mn$^{2+}$ ion nuclear charge and the charge of completely filled atomic shells. The expectation value $(\sum_{ms_z}\hat n_{ms_z}-Z_{Mn})$ is zero for an isoelectronic substitutional impurity. \cite{Elliot86} The short-range difference between these potentials contributes to the potential scattering $\hat H_{scat}$ and energies of the {\it 3d} and band states.

Similarly terms
\begin{equation}
\sum_{m s_z, m's'_z,q}(V_{ms_z,m's'_z,q,m's'_z} \hat n_{m's'_z}d^+_{ms_z}a_q+h.c.)
\end{equation}
and
\begin{equation}
\sum_{\mu,\nu, s_z,j_z} (U_{m s_z,\nu s_z}d^+_{m s_z}c_{\nu s_z}+U_{m s_z,\mu j_z} d^+_{m s_z}b_{\mu j_z}+h.c.)
\end{equation}
as well as
\begin{equation}
\sum_{m s_z,q,p}(V_{p,ms_z,p,q} \hat n_p d^+_{ms_z}a_q+h.c.)
\label{Vpq}
\end{equation}
are included in  $\hat H_{hyb}$. The term in Eq.(\ref{Vpq}) tends to compensate the effect of crystal ions potential on the {\it sp-d} mixing. \cite{Elliot86}

Similarly the term
\begin{equation}
\sum_{ms_z,m's'_z,p} V_{p,ms_z,p,m's'_z}\hat n_{p}d^+_{ms_z}d_{m's'_z}
\end{equation}
tends to compensate  the contribution of $\sum_{ms_z,ms'_z} U_{ms_z,m's'_z} d^+_{ms_z}d_{m's'_z}$ and the crystal ions core potential to energies of {\it 3d} and band states and the potential scattering of {\it 3d}-states.

By taking  into account the Coulomb interaction  the Hamiltonian in Eq.(\ref{Hamiltonian}) is transformed to
\begin{equation}
\hat H=\hat H_0+\hat H_{scat}+\hat U_0,
\label{Hamiltonian1}
\end{equation}
where $\hat U_0$ contains terms generated by $V_{sc}+H_{imp}$ that are not included in $\hat H_0+\hat H_{scat}$. They are responsible for various processes involving one, two or three {\it 3d}-electrons and include those responsible for the nonradiative recombination.

The Hamiltonian in Eq.(\ref{Hamiltonian1}) can be rewritten in the familiar form: \cite{Schriffer67,Hirst70}

\begin{equation}
\hat H=\hat H_{QD}+H_{Mn}+\hat H_{int},
\end{equation}

where $\hat H_{int}$ is the interaction between the QD electron system and the Mn$^{2+}$ ion.
The hybridization term $\hat H_{hyb}$ in Eq.(\ref{Hamiltonian1}) can be replaced by an effective scattering Hamiltonian, \cite{Bhatt92,Hirst71} so that $\hat H$ can be rewritten as

\begin{equation}
\hat H=\hat H_{QD}+\hat H_{Mn}+\hat H_{scat}+\hat U_0+\sum_{I}\frac{\hat H_{hyb}|I\rangle\langle I |\hat H_{hyb}}{E_i-E_I},
\label{Ham}
\end{equation}

where $E_i$ and $E_{I}$ are energies of the initial $|i\rangle$ and intermediate $|I\rangle$ states, respectively.

In addition to the terms $\hat H_{c~eff}$ and $\hat H_{v~eff}$ describing scattering of conduction and valence electrons on the {\it 3d}-shell the effective Hamiltonian contains the term $\hat H_{mix}$ responsible for the mixed scattering.

As it is shown in Ref.\cite{Bhatt92,Hirst71} the Hamiltonian
\begin{equation}
\hat H_{v~eff}=\sum_{I}\frac{\hat H_{pd}|I\rangle\langle I |\hat H_{pd}}{E_i-E_I}
\end{equation}
contains terms responsible for the potential scattering of valence electrons, contribution into energy of the {\it 3d}-level and terms of the {\it p-d} kinetic exchange that can be expressed in the Heisenberg form similar to $\hat H_{x}$ in Eq.(\ref{exchange}).\cite{Bhatt92} Analogous terms are contained in $\hat H_{c~eff}$.

The introduction of effective Hamiltonian $\hat H_{eff}$ is justified in the  ionic limit of the Anderson model. \cite{Hirst70} It takes place  when the broadening of the single electron {\it 3d}-states due to their hybridization with delocalized band states (QD states do not contribute to the broadening) is smaller than the energy gaps between the ground $d^5$ and excited $d^4$, $d^6$ configurations. \cite{Hirst70} In this limit the ground and excited states of Mn$^{2+}$ ions are characterized by certain spins $S=5/2$ or 3/2, which is obviously the case here.

The initial state, which we denote as [$d^5eh$], contains the Mn$^{2+}$ ion $d^5$ in the ground states  and the $e-h$ pair. Its energy is $E_i=E_0(5)-E_v+E_c$, where $E_0(n)$ is the energy of $d^n$ configuration of the Mn$^{2+}$ ion, $E_v$ and $E_c$ are energies of electron states at the bottom of conduction and the top of the valence bands. Intermediate states $|I\rangle$ are of two kinds: $|I^+\rangle$ corresponding to the virtual absorption  and $|I^-\rangle$  corresponding to the virtual emission of one electron. The dominant contribution to energy transfer comes from the virtual levels with minimal energy gaps $E_i-E^{\pm}$.  The energies of those states [$d^6h$] and [$d^4e^2h$] are  $E^+=E_0(6)+U_{eff}-E_v$, and $E^-=E_0(4)+2E_c-E_v$, respectively. The values of energy gaps $E_i-E^{\pm}$  are found in Appendix B.

By using commutational relations and the fact that virtual levels involve only $d^6$ and $d^4$ configurations the term
\begin{equation}
\hat H_{mix}= \sum_I\left( \frac{\hat H_{pd}|I\rangle\langle I|\hat H_{sd} +\hat H_{sd}|I\rangle\langle I | \hat H_{pd}}{E_i-E_I}\right)
\end{equation}
may be expressed as \cite{Hirst71}
\begin{equation}
\hat H_{mix}=\sum_{\mu \nu ,ms_z,m's'_z,j_z} (K_{mix}d^+_{ms_z}d_{m's'_z}b^+_{\nu j_z}c_{\mu s_z}
\label{I}
\end{equation}
$$
-\sum_{\mu \nu ,ms_z,j_z}K^+_{ms_z ms_z\nu \mu j_z}b^+_{\nu j_z}c_{\mu s_z}+h.c.),
$$
where coefficients are
\begin{equation}
K^{\pm}_{ms_z m' s'_z\nu \mu j_z}=-V_{sd~ ms_z\mu s_z}V^*_{pd~ m's'_z\nu j_z}\frac{1}{E_i-E^{\pm}}
\label{Mif1}
\end{equation}
and $K_{mix}=K^+_{ms_z m' s'_z\nu \mu j_z}+K^-_{ms_z m' s'_z\nu \mu j_z}$. The second term in Eq.(\ref{I}) leads to the interband potential scattering whereas the first one contributes to the nonradiative recombination as it is shown in the following section.

The Hamiltonian given by Eq.(\ref{Ham})  can be generalized to the case of many non-interacting Mn ions  which corresponds to  the limit of small Mn content when the interaction between Mn ions can be omitted.  The antiferromagnetic Mn-Mn coupling appears within the Anderson model in the fourth order of perturbation series and becomes important starting from  $x\simeq 0.05$.  The Mn-Mn interaction leads to the formation of Mn-Mn pairs, triads and clusters.  The antiferromagnetic Mn-Mn interaction leading to the cluster formation affects the dynamics of energy migration between Mn ions. The influence of pairs and triads on energy transfer is considered in Sec V. The discussion of the dynamics of Mn-Mn energy transfer is out of the scope of current manuscript.

Thus, $\hat U_0+\hat H_{mix}$ can be considered as a perturbation of the Hamiltonian
\begin{equation}
\hat H'=\hat H_{QD}+\hat H_{Mn}+\hat H_{ex}+\hat H_x,
\end{equation}
where $\hat H_{ex}$ is the {\it sp-d} kinetic exchange term. The potential scattering is dropped as it does not directly contribute to the nonradiative recombination.

The Coulomb and direct exchange interactions leading to the nonradiative recombination are contained in $\hat U_0$. In the following section we consider contributions of $\hat U_0$ and $\hat H_{mix}$ to the recombination separately and estimate relative efficiencies of the Coulomb and {\it sp-d} mechanisms.

\section{Mechanisms of nonradiative energy transfer to 3d-shell}

\subsection{Nonradiative  recombination due to Coulomb interaction}

The nonradiative recombination due to the Coulomb interaction is usually considered as a transition of the first order. \cite{Robbins,Allen86,Yassievich93} The only term in $\hat U_{0}$ leading to such a transition is
\begin{equation}
\hat V_{t}=\sum_{\mu \nu s'_z j_z s_z}\sum_{m, m'}^5 (V_{m,
\mu, \nu, m'}d_{ms_z}^{+}c_{\mu s'_z}^{+}d_{m' s_z}b_{\nu j_z}+$$
$$
V_{m,
\mu,  m',\nu
}d_{m s_z}^{+}c^+_{\mu s'_z} b_{\nu j_z}d_{m' s_z}+h.c.),
\label{Cou}
\end{equation}
where $V_{m,\mu, m',\nu}=\langle d_{ms_z},\mu s'_z |v_{sc}| d_{m's_z},\nu j_z\rangle$ and $V_{m,\mu, \nu, m'}=\langle d_{ms_z},\mu s'_z |v_{sc}|\nu j_z, d_{m's_z}\rangle$.

Contrary to the potential scattering where the single-electron term $H_{imp}$ tends to compensate the effect of {\it sp-d} Coulomb potential, $H_{imp}$ does not participate in the nonradiative recombination because matrix elements between initial and final states including different {\it 3d}-configurations vanishes.

The rate of resonant energy transfer from the photoexcited QD into Mn$^{2+}$ {\it 3d}-shell is given by the Fermi's golden rule:
\begin{equation}
R_{nr}=\frac{1}{\tau_{nr}}=\frac{2\pi}{\hbar}\frac{1}{N_i}\sum_{if}|M_{fi}|^2\delta(E_i-E_f),
\end{equation}
where $N_i$ is the number of initial states. \cite{Yassievich07}

The initial state of the system  $|i(5/2,S_z,G)\rangle=^6\hat A_1(5/2,S_z)\hat \psi_{ex}(G)|0\rangle$ consists of 5 {\it 3d}-electrons and the {\it e-h} pair while the final state $|f(3/2,S'_z)\rangle=^4 \hat T_1(3/2,S_z)|0\rangle$ is the excited state of the Mn$^{2+}$ ion. Here $\psi_{ex}(G)=\hat \psi_{ex}(G)|0\rangle$ is the exciton state, where $G$ denotes a set of quantum numbers characterizing it.  In the absence of magnetic field $G$ is a representation of the symmetry group of exciton state. For instance, it is $E$ for bright and $A_1, A_2$ for dark excitons in quantum dots and wells of $D_{2d}$ symmetry. The exciton creation operator $\hat \psi_{ex}(G)$ is expressed via $c_{\mu s_z}^{+}$ and $b_{\nu j_z}$ operators. In strong magnetic field ${\bf B}\parallel {\bf 0z}$ exciton states are characterized by $J_z$. Operators $^4\hat T_1(3/2,S_z)$ and $^6\hat A_1(5/2,S_z)$ are sums of products of five $\hat d^+_{ms_z}$ operators. Wave-functions of the many-electron $^4 T_1(3/2,S_z)=^4\hat T_1(3/2,S_z)|0\rangle$ and $^6A_1(5/2,S_z)=^6\hat A_1(5/2,S_z)|0\rangle$ states in various approximations of the crystal field theory can be found in literature. \cite{PRB05,Schriffer67,Griffith,Bhatt92, Blinowski91}

The matrix element of nonradiative recombination $M_{fi~S'_z,S_zG}=\langle f(3/2,S'_z)|\hat V_t|i(5/2,S_z,G)\rangle$ reduces to a sum of Coulomb and exchange integrals. The Coulomb integrals lead to dipole-dipole energy transfer from QDs to Mn$^{2+}$ ions as it is shown in Appendix A.  The   anticommutation between $3d$ and band electrons does not affect Coulomb matrix elements so that energy transfer due to the Coulomb interaction can be understood as a result of emission and absorption of virtual photons by QD states and Mn$^{2+}$ ions. The crystal media can screen this process but cannot interfere with it. \cite{Andrews04}  The dipole transition between $^4T_1(3/2)$ and $^6A_1(5/2)$ states is spin-forbidden and, therefore, admixing of Mn$^{2+}$ excited states with spin 3/2 to  $^6A_1(5/2)$ is required.

Robbins and Dean considered the dipole-dipole mechanism as the dominant in nonradiative exciton decay accompanied by intraionic excitations. \cite{Robbins}  Similar point of view is expressed in Ref.\cite{Yassievich93,Yassievich07, Yassievich97} where excitation of  $Er$ ions in $Si$ accompanied by optically forbidden interconfigurational transitions of $Er$ is considered. However, efficiency of  the exchange mechanism can substantially exceed that of the Coulomb one when the latter is spin-forbidden, which is usual in atomic and molecular systems. \cite{Agranovich,Photochem} Analyzing the process of deexcitation of Mn$^{2+}$ ions in the presence of free carries, in other words, the reverse process in respect to the impact excitation (Auger process),  Allen concluded that the Coulomb interaction underestimated it as much as by two orders of magnitude and, therefore, the main mechanism of energy transfer was related to the exchange interaction rather than the Coulomb one. \cite{Allen86} The idea about domination of exchange mechanism in excitation of Mn$^{2+}$ ions is widely accepted now. \cite{Nawrocki95, Gamelin,Abramishvili91} The exchange mechanism of energy transfer in atomic and molecular systems is subject to the Wigner spin conservation rule, which states that it is allowed if the total spin of the interacting system is conserved. \cite{Photochem}

The spin conservation rule was used by Nawrocki {\it et al.} to derive selection rules for the exchange mechanism. \cite{Nawrocki95} However, in order to apply it the authors of Ref.\cite{Nawrocki95} simplified the studied system so that they neglected any spin-orbit coupling in the system and consider the transition $ |^6A_1(5/2)\varphi^e_{s_z} \rangle\to | ^4T_1(3/2)\varphi^{e}_{s'_z}\rangle$ instead of exciton recombination. Exciton states in QDs are not completely characterized by spin due to the strong spin-orbit interaction in the valence band and {\it e-h} exchange interaction. However, spin selection rules for the exchange mechanism can be derived from the analysis of exciton and Mn$^{2+}$ spin functions.

The matrix element of the nonradiative recombination due to Coulomb interaction in the first-quantized form is $M_{fi}=\langle f|V_{sc}|i\rangle$, where $|i\rangle=|\hat A(^6A_1(5/2,S_z)\psi_{ex})\rangle$, $|f\rangle=|\hat A (^4T_1(3/2 S'_z)\psi_0)$, $\psi_{ex}$ is the exciton many-electron function, $\psi_0$ is the ground state of the crystal, and $\hat A$ is  the antisymmetrization operator.

The exciton states are sums of products of spacial and spin functions corresponding to spins $s_{ex}=0,1$ and spin projections $s_{ex~z}=0$ for bright and $s_{ex~z}=\pm 1$ for dark excitons because the wave-functions of bright exciton ground state can be expressed as $\psi^b_{ex}(r_1,r_2)=F_{hh0}(r_1)F_{e0}(r_2)/\sqrt{2}(-(X+iY)_1S_2\alpha_1\beta_2\pm (X-iY)_1S_2\beta_1\alpha_2)=-F_{hh0}(r_1)F_{e0}(r_2)/\sqrt{2}(X_1S_2(\alpha_1\beta_2\mp \beta_1\alpha_2)+iY_1S_2(\alpha_1\beta_2\pm \beta_1\alpha_2))$ whereas the dark exciton wave-functions are $\psi^d_{ex}( r_1, r_2)=-F_{hh0}(r_1)F_{e0}(r_2)/\sqrt{2}(X+iY)_1S_2\alpha_1\alpha_2$, and $\psi^d_{ex}(r_1, r_2)=F_{hh0}(r_1)F_{e0}(r_2)/\sqrt{2}(X+iY)_1S_2\beta_1\beta_2$. Here $\alpha$ and $\beta$ are $s_z=1/2$ and $s_z=-1/2$ spin states, respectively.

Unlike the Coulomb processes the antisymmetrization between {\it 3d-} and band electrons is crucial for the exchange mechanism. The wave-function of the final state $|f\rangle$ is that of the crystal in the ground state $\psi_0$ containing one Mn$^{2+}$ ion in the excited state $^4T_1(3/2,S_z)$. The final state is a sum of Slater determinants constructed from $N+5$ single-electron wave-functions where $N$ is the number of valence and conduction electrons of the crystal including two Mn {\it 4s} electrons that become delocalized. Each Slater determinant can be presented as a sum of products of spatial and spin functions, where spin functions are basis functions of the irreducible representation of the permutation group of $N+5$ electron spins corresponding to certain squared spin $\hat S^2$ and spin projection $S_z$, whereas spatial functions are basis functions of the representation conjugate to the spin representation. \cite{Elliots}  It means that the final state certainly contains spin functions with $S=3/2$.

The initial state $|i\rangle$ obviously contains Slater determinants with spin projections $S_z=\pm 5/2$ and $S_z=\pm 7/2$, which means that corresponding spin functions have S=5/2 and 7/2. They can also have  $5/2-1=3/2$ contained in $5/2\pm s_{ex}$. It means that the nonradiative transition is allowed for bright and dark excitons. Similarly it is allowed for $X^-$ trions. When the exchange mechanism is allowed it necessarily leads to the conservation of the spin-projection $S_z+s_{ex~z}$ because of properties of the exchange integral $V_{m\mu \nu m'}$.

The exchange matrix elements do not allow parametrization and hardly be estimated in a simple manner. They can be small or large because the {\it s-d} exchange constant is about a quarter of the magnitude of the {\it p-d} kinetic exchange constant, \cite{Larson88} whereas contribution of the direct exchange into the {\it p-d} exchange is usually assumed to be zero. \cite{Larson88, Merkulov99}

The use of the Fermi golden rule is correct if the matrix elements of perturbations $\hat H_{mix}$ and $\hat V_t$ are much smaller than energy difference between any Mn$^{2+}$ configurations and {\it e-h} QD states. It is valid for the {\it sp-d} and Coulomb mechanisms as it is shown in Appendices. It is natural to expect that this is also valid for the exchange mechanism.

The hybridization $\hat H_{hyb}$ makes it possible Coulomb processes via virtual states involving $d^4$ and $d^6$ configurations because of the second-order term
\begin{equation}
\sum_I \frac{\langle i|U_0|I\rangle\langle I|H_{hyb}|f\rangle}{E_i-E_I}+\frac{\langle i| H_{hyb}|I\rangle\langle I|U_0|f\rangle}{E_i-E_I}
\end{equation}

Its contribution to energy transfer is small first of all because matrix elements $H_{hyb~If}/(E_i-E_I)$ and $H_{hyb~iI}/(E_i-E_I)$ are much smaller than unity as it is shown in Appendix B. Besides, the Coulomb matrix elements between $|i\rangle$ $|f\rangle$ and intermediate levels $|I\rangle$ involving  $d^4$ and $d^6$ configurations with spin $S=2$ is spin-forbidden.

\subsection{Nonradiative recombination due to sp-d mixing}

The {\it sp-d} mixing is responsible for the nonradiative recombination that can be understood as a result of successive hopping of the electron(hole) and hole(electron) in the {\it 3d}-shell. The hopping of a valence electron in and out of the {\it 3d}-shell is the reason for the kinetic {\it p-d} exchange interaction. The hopping of conduction electrons becomes possible because of the admixture of valence band states to conduction ones as ${\bf k}\neq 0$. It is responsible for the reduction of the {\it s-d} exchange interaction reported in Ref.\cite{Merkulov99} The {\it sp-d} mechanism was proposed by Schmitt-Rink  {\it et al.} as a mechanism of excitation of rare-earth ions incorporated into a semiconductor matrix. \cite{SchmittRink91} It is related to the term
\begin{equation}
\hat H_{mix}=\sum_{m, m',\nu,\mu,s_z,s'_z,j_z} (K_{mix}d^+_{ms_z}d_{m's'_z}b^+_{\nu j_z}c_{\mu s_z} +h.c.),
\end{equation}
The {\it p-d} coupling coefficient for the valence electron ground state, $\nu=0$, is $V_{pd~ms_z0j_z=\pm 3/2}=\langle d_{ms_z}|\hat U_{cr}|\varphi^v_{0 \pm 3/2}\rangle=\mp F_{hh0}({\bf a}_0)/\sqrt{2}\langle d_{ms_z}|\hat U_{cr}|(X\pm iY)_{\pm 1/2}\rangle$, where ${\bf a}_0$ is the position of Mn$^{2+}$ ion. The {\it s-d} coupling coefficient for the conduction electron ground state $\mu=0$ is  $V_{sd~ms'_z,0s_z}=\langle d_{ms'_z}| \hat U_{cr}|\varphi^e_{0~s_z}\rangle= F_{ex0}({\bf a}_0)\langle d_{m s'_z}|\hat U_{cr}|X_{s_z}\rangle+F_{ey0}({\bf a}_0)\langle d_{ms'_z}|\hat U_{cr}|Y_{s_z}\rangle+F_{ez0}({\bf a}_0)\langle d_{m s'_z}|\hat U_{cr}|Z_{s_z}\rangle$ since $\langle d_{m s'_z}| \hat U_{cr}|S_{s_z}\rangle=0$ because of the symmetry. \cite{Larson88,Hass,Bhatt92,Blinowski91}

Due to strong anisotropy of QDs under study the main contribution to $V_{sd}$ is because of quantization along {\bf 0z}. The coefficient of {\it s-d} mixing for the electron ground state, $\nu=0$, in a QD can be expressed as:
\begin{equation}
V_{sd~ms'_z,0s_z}= F_{ez0}({\bf a}_0)\langle d_m|\hat U_{cr}|Z\rangle\delta_{s'_z,s_z}=
\label{Vsd}
\end{equation}
$$
F_{ez0}({\bf a}_0)V_{pd~mz}\delta_{s'_z,s_z}
$$
at ${\bf k}$ close to the center of Brillouin zone. Coefficients  $V_{pd~ms_z0j_z}$ and $V_{sd~ms_z0s_z}$ are related to the {\it p-d} hopping amplitude $V^0_{pd}$ as it is shown in Appendix B.

The matrix element of recombination of the bright exciton $|^6A_1(S_z)J_z=\pm 1\rangle \to |^4T_1(S'_z)\rangle$ calculated in Appendix B is:
\begin{equation}
M_{mix}=
\alpha(S_z)\frac{16}{N_0}F_{hh0}({\bf a}_0 )F_{ez0}({\bf a}_0)|V^0_{pd}|^2
\label{Mif0}
\end{equation}
$$
\times\left(\frac{1}{U_{eff}+\epsilon_d-E_c}+\frac{1}{E_v-\epsilon_d}\right)\left(\frac{\gamma}{2}\right)^{1/2}\delta_{S'_z,S_z},
$$
where $\alpha(S_z)$ are certain coefficients found in the Appendix B. The factor $\sqrt{\gamma/2}$ appears because of admixing  of valence band states to conduction ones.  It is important that such a transition conserves the Mn$^{2+}$ ion spin projection $S_z=S'_z$.

The {\it p-d} exchange constant appearing in the {\it hh} exchange Hamiltonian $\hat H_{ex}=-\beta/3 {\bf jS}({\bf a}_0)\delta({\bf r}-{\bf a}_0)$ is given by the expression: \cite{Larson88,Hass,Bhatt92}
\begin{equation}
\beta=-\frac{32}{5}\frac{1}{N_0}|V^0_{pd}|^2\left(\frac{1}{U_{eff}+\epsilon_d-E_v}+\frac{1}{E_v-\epsilon_d}\right)
\label{beta}
\end{equation}

This expression is obtained neglecting the crystal field splitting of Mn$^{2+}$ states. \cite{Larson88,Schriffer67} This result is reproduced with due regard for crystal field splitting of  the initial state. \cite{Bhatt92}  It is natural to expect that neglecting the crystal field splitting in matrix element $M_{mix}$ in Eq.(\ref{Mif0}) also  does not noticeably affect the result. The physical meaning of {\it sp-d} mechanism is illustrated in Fig. 1. The process of kinetic {\it p-d} exchange interaction is shown in Fig.2 for comparison.

The {\it sp-d} mechanism of nonradiative recombination can be much more effective than that due to the Coulomb interaction. Estimates presented in Appendices A and B provide the ratio of recombination rates of the Coulomb and {\it sp-d} mechanisms for CdMnSe/ZnSe QDs averaged over distribution of Mn ions  $\overline R_{Coul}/\overline R_{mix}\simeq 10^{-2}$. The envelope $F_{es}({\bf r})$ reaches maximum at the center of QDs, whereas $F_{ez}({\bf r})$ reaches maximum at the QDs boundaries, \cite{Merkulov00}  therefore the ratio $R_{Coul}/R_{mix}$ depends on the distribution of Mn ions within the {\it e-h} pair localization volume. The ratio $R_{Coul}/R_{mix}$ is expected to be larger than $\overline R_{Coul}/\overline R_{mix}$ in CdMnSe/ZnSe  and smaller in CdSe/ZnMnSe QDs.

Although the {\it sp-d} mechanism was proposed for excitation of rare-earth ions in a semiconductor \cite{SchmittRink91} it better matches our case. For instance, {\it 4f}-orbitals of Er ions in Si are located  much below the top of valence band ($\sim$ 10 eV), \cite{Yassievich97} which results in weak {\it p-f}  and {\it s-f} hybridizations. Besides 4f-orbitals are closely located to the ion core so that their interaction with bands electrons is weak. In contrast, the {\it 3d}-level lays at 3.4 eV below the top of valence band \cite{Larson88,Hass} and resonantly mixes with valence band states.

The dipole-dipole mechanism is an analogue of the Forster mechanism of energy transfer between atoms or molecules whereas the exchange mediated mechanism is similar to the Dexter one. \cite{Photochem}  The mechanism related to the {\it sp-d} mixing can also be associated with the Dexter mechanism, where $\hat H_{mix}$ plays a role of the effective exchange interaction.

The contribution of the direct exchange mechanism to the nonradiative recombination can hardly be estimated in a simple manner so that relative contributions of $M_{ex}$ and $M_{mix}$ is unknown. However, both  mechanisms lead to spin-dependent energy transfer. Possible manifestation of such a process is discussed in Sec. V.

It is worth noting  that the {\it sp-d} mechanism can contribute to the impact excitation of Mn$^{2+}$ ions. The excitation  via either intra- or interband electronic transitions is possible. \cite{Yassievich93,SchmittRink91} The impact excitation by means of the intraband transition becomes allowed when the energy of an optically or electrically excited electron with respect to the bottom of conduction band exceeds 2.15 eV. The impact excitation can be understood as a result of capture of such an excited conduction electron on the {\it 3d}-shell because of the {\it s-d} mixing with successive escape of one of {\it 3d}-electrons with the opposite spin also into the conduction band.   Ayling and Allen state, however, that the efficiency of excitation of Mn$^{2+}$ via interband transitions is much larger than that via intraband ones. \cite{Allen86,Allen87}

\section{Phonon-assisted processes and energy transfer into upper Mn$^{2+}$ excited states}

The resonant processes discussed above assume equality of the {\it e-h} pair transition energy and the energy of intraionic excitation  $E_0\simeq 2.15$ eV. Experiments show that the photoluminescence quenches even when transition energy of the ground state of {\it e-h} pairs in QDs substantially exceeds $E_0$. \cite{PRB05,Henneberger01}  There are two mechanisms, which allow such processes: the phonon-assisted recombination and the nonradiative transition in upper Mn$^{2+}$ states.

The model of phonon-assisted energy transfer from a semiconductor crystal to rare-earth impurities is proposed by Yassievich {\it et al.} \cite{Yassievich93} It is based on the single configurational coordinate scheme describing the electron-phonon interaction of  impurity electrons with a phonon mode of the crystal.  According to the model  nonresonant recombination is possible due to emission of multiple phonons. The probability of the phonon-assisted transition is given by:
\begin{equation}
R^{pn}_{nr}=\frac{2\pi}{\hbar}\frac{1}{N_i}\sum_{i
f}\sum_{N_{ph}}|M_{fi}|^2I(N_{ph})\delta(E_g-E_0-N_{ph}\hbar\omega_{ph}),
\label{phonon}
\end{equation}
where $N_{ph}=(E_g-E_0)/\hbar\omega_{ph}$. At low  temperatures ,$k_BT\ll \hbar\omega_{ph}$, the coefficient $I(N_{ph})\simeq (S_H)^{N_{ph}}e^{-N_{ph}}/N_{ph}!$, where $S_H$ is the Huang-Rhys factor, which determines the strength of electron-phonon coupling. \cite{Abakumov}  According to the recipe of Ref.\cite{Yassievich93} summation over $N_{ph}$ is replaced by integration and leads to the expression:
\begin{equation}
R^{pn}_{nr}(K_{fi})=\frac{1}{\tau_{nr}}=\frac{2\pi}{\hbar}\frac{I(K_{fi})}{\hbar
\omega_{ph}}\frac{1}{N_i}\sum_{i f}|M_{fi}|^2\label{prob}
\end{equation}

The probability $R^{pn}_{nr}(K_{fi})$ has the maximum at $K_{fi}=S_H$ at large coupling strength $S_H>1$ and quickly decreases with $K_{fi}$.  The upper limit of the efficiency of the phonon-assisted recombination can be estimated by the summation over all $K_{fi}$.

The single configurational coordinate picture is often used for the analysis of temperature dependence of PL lines of Mn$^{2+}$ internal transitions and the Stocks shift between Mn$^{2+}$ absorbtion and emission spectra. \cite{Kikoin,Gamelin,Galazka} The structures under study are characterized by more than 200 meV wide  Mn$^{2+}$ PL line \cite{PRB05} that can be attributed to strong interaction  of the {\it 3d}-shell with crystal vibrations ($S_H\gg 1$). Assuming that the energy of the vibrational mode is of order $\hbar\omega_{LO}$ one obtains from (\ref{prob}) $\tau_{nr}\simeq  10^{-11}$ s for a Cd$_{0.85}$Mn$_{0.15}$Se/ZnSe QDs with reasonable values of $D\simeq 35$\AA~and $L\simeq 12$\AA~.

Although this value of $\tau_{nr}$ is not far from measured PL decay times $\tau_0=20-80$ ps of CdMnSe/ZnSe QDs  with close parameters at $B=0$ T, \cite{Toropov} small intensities of the longitudinal optical phonon (LO-phonon) replicas of Mn$^{2+}$ PL lines in DMS materials indicate weak coupling between the {\it 3d}-shell and LO-phonons. \cite{Agekyan} Similarly these structures are characterized by a weak vibronic interaction. \cite{Kikoin} In this case recombination processes into upper excited Mn$^{2+}$ states can dominate because nonradiative energy transfer is possible not only into the lower excited state $^4T_1$ but also into other excited states $^4T_2,^4E_1,^4A_1$ located 0.2-0.7 eV above $^4T_1$. The large inhomogeneous broadening of Mn$^{2+}$ PL lines makes nonradiative transitions possible in a wide range of energies higher than 2.15 eV. The excited $^4T_1, ^4T_2, ^4A_1, ^4E$ states have spins $S=3/2$, so that relaxation into $^4T_1$ state is expected to be very fast.

\section{Spin-dependent selection rules and increase in PL intensity in magnetic field.}

The strong increase in PL intensity of an ensemble of Cd(Mn)Se/Zn(Mn)Se QDs in the magnetic field $B\parallel 0z$ is explained as a result of suppression of the exciton nonradiative recombination  because of depopulation of $A_1(5/2,S_z)$ states with $S_z> -5/2$. \cite{Kim98,Nawrocki95} The results of Sec. III  reveal that the Coulomb process is allowed for $S_z=\pm 5/2$ and it does not contribute to the increase in PL intensity.

The exchange and  {\it sp-d} mechanisms lead to conservation of the total spin projection  $S_{z}+s_{ex,z}=const$. Recombination of bright excitons ($s_{ex,z}=0$) requires $\Delta S_{z}=0$, whereas the recombination of dark excitons ($s_{ex,z}=\pm 1$) is possible when $\Delta S_{z}=\mp 1$.

The selection rules predict that the nonradiative recombination of bright excitons involving $S_z= 5/2$ is spin-forbidden, whereas it is allowed for the dark exciton with $J_z=2$($s_{ex,z}=1$) and forbidden for that with $J_z=-2$($s_{ex,z}=-1$). The strong increase in PL intensity requires slow relaxation of bright excitons to $J_z=2$ dark states. Phonon-assisted recombination and energy transfer into upper excited Mn$^{2+}$ levels do not break the rules.

However, the selection rules are not valid for {\it e-h} complexes formed by hole states with large admixture of {\it lh}-states, which is the case for bulk DMS materials of cubic symmetry. In contrast,  the selection rules should be valid for highly anisotropic crystals of wurtzite structure such as CdMnS, which is in agreement with the results of photoreflectance measurements reported by Nawrocki {\it et al.} \cite{Nawrocki95}

Recent studies of the optically detected magnetic resonance give additional arguments in support of spin-dependent energy transfer from ensemble of CdMnSe/ZnSe QDs to Mn$^{2+}$ ions and its dependence on the direction of magnetic field. \cite{Baranov}

The observed increase in QD PL intensity indicates that spin-dependent {\it sp-d} and exchange mechanisms dominate over the  Coulomb one. This assumption correctly explains magnetic field and temperature dependence of the PL intensity of QD ensemble. The following empirical expression is found to fit the experimentally observed increase in PL intensity with magnetic field ${\bf B}\parallel {\bf 0z}$: $I(B)=A/(1+Cp(B/T))$, where $p(B/T)\simeq \alpha+\beta e^{-\Delta E(B)/{k_BT}}$, $\alpha$ and $\beta$, $A$ and $C$ are constants. \cite{PRB05,Kim98} This well-known expression describes the temperature dependence of the interband PL intensity in the presence of nonradiative recombination centers. The magnetic field dependent activation energy is $\Delta E(B)=\Delta_{eh}+\Delta_{Mn}$, where $\Delta_{eh}=\mu_B(g_e+g_h)B$ is the energy of Zeeman splitting of {\it e-h} states.  The parameter $\alpha$ is the probability of nonradiative recombination independent of magnetic field while $\beta$ is the probability of nonradiative recombination involving the $S_z=-3/2$ level. \cite{PRB05,Kim98,Nawrocki95} Results of Ref.\cite{PRB05} indicate that the ratio $I(B)/I(0)\simeq (1+\beta/\alpha)$ can reach $10^2$ in high magnetic fields  $\Delta E(B)/{k_BT} \gg 1$, which is possible if $\beta\ll \alpha$ and the relaxation of the bright excitons into lower dark states is slow.

Since the radiative life-time of $^4T_1(3/2)\to ^6A_1(5/2)$ transition is substantially longer than the time of  nonradiative recombination, fast saturation of $I(B)$ is expected because only a few tens of Mn$^{2+}$ ions can interact with the localized {\it e-h} pair. The derivation of $I(B)$ implies, however, that the reservoir of Mn$^{2+}$ ions is infinite. This fact can be related to the fast energy diffusion within ensemble of Mn$^{2+}$ ions. \cite{Agekyan} The dependence $I(B)$ very well describes experimental results, \cite{PRB05,Lee05,Kim98} which supports the assumption.

The QDs considered in Ref.\cite{PRB05,Kim98} are assumed to be neutral meanwhile considerable amount of QDs in the ensemble can be negatively charged due to a n-type background doping almost always present in II-VI materials. The selection rules predict that $X^-$ $j_z=\pm 3/2$ states are involved in nonradiative recombination at $B=0$. In magnetic field ${\bf B}\parallel {\bf 0z}$, which polarizes both trion and Mn$^{2+}$ states only $j_z=+3/2$ trion participates in the nonradiative recombination whereas $j_z=-3/2$ trion does not.

At Mn content of $x=1-2\%$ the lower $X^-$ trion state in magnetic field  ${\bf B}\parallel {\bf 0z}$ is that with $j_z=3/2$. \cite{JETP07} Negatively charged QDs do not lead to the increase in PL intensity when  the relaxation of the $j_z=-3/2$ to the $j_z=3/2$ state is fast.

However, measurements of $X^-$ trions in individual CdSe/ZnSe/ZnMnSe QDs reported in Ref.\cite{JETP07} show surprising increase in the intensity of upper $\sigma^-$ component of the trion PL in magnetic field ${\bf B}\parallel {\bf 0z}$. This result can be understood on the ground of the proposed model if the relaxation $j_z=-3/2$ to $j_z=3/2$ is slow.  The two level model, which takes into account the nonradiative recombination, relates the time of nonradiative recombination $\tau_{nr}$ to the time of spin relaxation $\tau_s$ from the upper trion state: $\tau_{nr}(B)<\tau_s/2(1-e^{-\Delta_{Mn}/kT})$, where  $\Delta_{Mn}=\mu_Bg_{Mn}B$ is the Zeeman splitting of nearest Mn levels. This is probably because the relaxation $-3/2\to 3/2$ requires changes $\Delta j_z=3$ so that it can be slow in QDs under study because of strong {\it lh-hh} splitting and the splitting of Mn$^{2+}$ spin states in magnetic field.  Negatively charged QDs, therefore,  can contribute to the increase in $I(B)$.

In the Voigt geometry (${\bf B}\perp {\bf 0z}$) magnetic field aligns Mn$^{2+}$ and electron spins opposite to the direction the field whereas {\it hh} hole moments remain directed along ${\bf 0z}$. In addition, all dark $|\pm 2\rangle$ and bright $|\pm 1\rangle$ exciton states mix at ${\bf B}\perp {\bf 0z}$. According to the selection rules the nonradiative recombination of dark excitons $J_z=2$ is allowed at any $B$, and, therefore, no increase in PL intensity is expected in the Voigt geometry, which is in agreement with the results presented in Ref.\cite{PRB05}  The nonradiative recombination of trions in the Voigt geometry is also allowed because of mixing of $\pm 3/2$ states.

The Mn-Mn interaction has been omitted in the foregoing discussion, meanwhile it causes formation of Mn clusters among which Mn-Mn  pairs and triads are most important. \cite{Gumlich85}  Number of the pairs quickly increases with increase in Mn content and reaches maximum at $x$ of several percents. The pairs are characterized by the relatively large antiferromagnetic coupling energy,  about 1 meV for nearest neighbors in CdMnSe and ZnMnSe, \cite{Larson88}  so that they remain coupled even at high magnetic field of about 12 T. Such pairs do not participate in the Zeeman splitting of band states as the total spin of the pair is zero. In contrast, the pairs can participate in radiative and nonradiative recombinations as those processes involve excitation of individual Mn$^{2+}$ ions in the pairs. The Mn-Mn coupling energy is still too weak to modify or mix $3d^5$ configurations of individual ions within the pair as the energy gap between {\it 3d}-configurations is several eV. Therefore, there are not many reasons to expect strong changes of either $R_{Coul}$ or the time of intraionic optical transition $1/\tau_r$. There are no reason to expect strong changes of $M_{mix}$ and $M_{ex}$ due to the pairs and triads formation.

Experimental results reported in Ref.\cite{Gumlich85} revealed, however, shorter PL life-time of {\it d-d} pairs line than that of individual Mn$^{2+}$ ions.  It can be assumed that the pair formation modifies Mn$^{2+}$ ion environment and changes covalent coupling strength, which, in turn, decreases $\tau_r$. Consequently, increase of $R_{Coul}$ and the ratio $\overline R_{Coul}/\overline R_{mix}$ with Mn$^{2+}$ content can be expected as $R_{mix}$ linearly depends on $x$.  The Mn-Mn pairs do not lead to the increase in  $I(B)$, because  magnetic field does not align spins of coupled Mn$^{2+}$ ions. Contrary to the pairs, Mn-Mn-Mn triads are characterized by nonzero average spin and therefore can contribute to the increase in $I(B)$.

The {\it sp-d} and exchange mechanisms are short-range ones, they are effective within the {\it e-h} pair localization volume. On the contrary, the dipole-dipole mechanism is a long-range one. It may be important in CdSe/ZnMnSe QDs and obscure spin-dependent effects. At distances  between the CdSe/ZnMnSe QD edge and the Mn$^{2+}$ ion core $R_d$ much larger than the QD dimensions $M_{Coul}$ decreases as $R_d^{-6}$. \cite{Andrews04,Agranovich}   Thus, the ratio $R_{Coul}/R_{mix}$ increases with $R_d$ and  the dipole-dipole mechanism can dominate at large distances. Assuming that the energy transfer between Mn ions is fast one can expect larger increase in $I(B)$ for  CdMnSe/ZnSe neutral QDs with $x<0.1$ and the smaller one for CdSe/ZnSe/ZnMnSe neutral QDs which is in agreement with experimental results. \cite{PRB05,Toropov} At large $x>0.1$ the fast saturation of $I(B)/I(0)$ or even decrease can take place.   This is not valid in general case because the ratio $I(B)/I(0)$ depends on various parameters such as  distribution of Mn$^{2+}$ ions, the ratio of neutral and charged QDs and probably considerably vary from sample to sample.

The considered mechanisms of nonradiative recombination are effective not only in DMS QDs but also in bulk and other low-dimensional structures. DMS II-VI materials are characterized by close parameters $U_{eff}$ and $E_v-\epsilon_d$  but slightly different $E_g$, $V_{sd}$ and $V_{pd}$. With the use of parameters reported in Ref.\cite{Larson88,Young91} one can find that the contribution of the {\it sp-d} mechanism to energy transfer in CdMnS and CdMnTe structures is comparable to that in CdMnSe ones.

The proposed model deals with direct energy transfer from {\it e-h} complexes to Mn$^{2+}$ ions. In bulk DMS materials and low-dimensional structures other processes via various intermediate states such as defects, impurities and surface states are possible. The mechanisms considered above are involved in energy transfer in those cases too.

\section{Conclusions}

Mechanisms of the nonradiative recombination of {\it e-h} complexes in Cd(Mn)Se/Zn(Mn)Se QDs accompanied by the intraionic excitation of Mn$^{2+}$ ions are analyzed within the framework of the single-electron model of deep {\it 3d}-level in semiconductors. Together with traditional mechanisms related to the Coulomb and exchange interactions between {\it 3d-} and band electrons another mechanism caused by the {\it sp-d} hybridization is considered. Estimates of matrix elements reveal that the efficiency of this mechanism can considerably exceed that of the Coulomb mechanism and its contribution to nonradiative recombination can be significant or even dominant. Mechanisms of energy transfer from neutral and negatively charged QDs to Mn$^{2+}$ ions due to the {\it sp-d} mixing and direct exchange interactions are subject to the spin selection rules $S_{z}+s_{ex~z}=const$ in magnetic fields ${\bf B}\parallel {\bf 0z}$, whereas the Coulomb mechanism does not obey them. These rules are because of the strong {\it hh-lh} splitting of hole states in Cd(Mn)Se/Zn(Mn)Se QDs.  It is shown that nonradiative recombination remains efficient even when the fundamental energy gap $E_g$ substantially exceeds the energy of the lower Mn$^{2+}$ internal transition most probably because of energy transfer into upper Mn$^{2+}$ excited states. The proposed model indicates that the increase in PL intensity depends on the Mn$^{2+}$ content, distribution of Mn ions, QD dimensions so that its magnitude can strongly vary from sample to sample.

\begin{acknowledgements}
This work has been supported by the Russian Foundation for Basic Research, FCT Project PTDS/FIS/72843/2006 and SANDiE Network of Excellence. The authors are grateful to I.~N.~Yassievich, S.~N.~Molotkov, V.~M.~Edel'stein for fruitful discussions and to S.~Diehm for reading the manuscript and supplying critical comments.
\end{acknowledgements}

\appendix

\section{ Matrix element of nonradiative recombination due to Coulomb interaction}

The only term in $\hat U_{0}$ leading to the transition from the initial $|i\rangle=^6\hat A_1(5/2,S_z)\hat \psi_{ex}(G)|0\rangle$ to final $|f\rangle=^4 \hat T_1(3/2,S_z)|0\rangle$ states in the first order is
\begin{equation}
\hat
V_{t}=\sum_{\mu \nu s'_z, j_z,s_z}\sum_{m, m'}^5 (
V_{m,
\mu,  m',\nu
}d_{m s_z}^{+}c^+_{\mu s'_z} b_{\nu j_z}d_{m' s_z}
\label{Couc}
\end{equation}
$$
+
V_{m,
\mu, \nu, m'}d_{ms_z}^{+}c_{\mu s'_z}^{+}d_{m' s_z}b_{\nu j_z}+h.c.),
$$

where $V_{m,\mu, m',\nu}=\langle d_{ms_z},\mu s'_z |v_{sc}| d_{m's_z},\nu j_z\rangle$ and
$V_{m,\mu, \nu, m'}=\langle d_{ms_z},\mu s'_z |v_{sc}|\nu j_z, d_{m's_z}\rangle$.

The Coulomb mechanism of nonradiative recombination originates from the first term in $\hat V_t$.
The screened Coulomb potential is
\begin{equation}
v_{sc}({\bf r}_2-{\bf r}_1)=\frac{e^2}{\epsilon_{\infty}|{\bf r}_2-{\bf r}_1|}=\frac{4\pi e^2}{\epsilon_{\infty}(2\pi)^3}\int \frac{e^{i{\bf q(r_2-r_1)}}}{q^2} d^3 q
\end{equation}

The dominant contribution to the matrix element of long-range Coulomb potential comes from the domain  $|{\bf r}_2-{\bf a}_0|\gg |{\bf r}_1-{\bf a}_0|\simeq a_B$. By using the multipole expansion of the Coulomb potential
\begin{equation}
v_{sc}({\bf r}_2-{\bf r}_1)=\frac{e^2}{\epsilon_{\infty}|({\bf r}_2-{\bf a}_0)-({\bf r}_1-{\bf a}_0)|}\simeq
\end{equation}
$$
\frac{e^2}{\epsilon_{\infty}|{\bf r}_2-{\bf a}_0|}+
\frac{({\bf r}_1-{\bf a}_0)({\bf r}_2-{\bf a}_0)}{\epsilon_{\infty}|{\bf r}_2-{\bf a}_0|^3}+...
$$
the matrix element $V_{m\mu m'\nu}$ is parameterized as
\begin{equation}
V_{m,\mu,m',\nu}\simeq \frac{e^2}{\epsilon_{\infty}} {\bf d}_{ms_zm's_z}{\bf P}_{\mu s'_z \nu j_z}=
\frac{e^2}{\epsilon_{\infty}}\langle d_{ms_z}|{\bf r}_1|d_{m's_z}\rangle
\end{equation}
$$
\times\langle \mu s'_z|\frac{({\bf r}_2-{\bf a}_0)}{|{\bf r}_2-{\bf a}_0|^3}|\nu j_z\rangle
$$
The Coulomb operator can be rewritten as $\hat V_{t~Coul}=\sum_{m m' s_z} {\bf d}_{ms_zm's_z}d_{ms_z}^{+}d_{m' s_z}$ $\sum_{\mu\nu s'_z j_z}{\bf P}_{\mu s'_z\nu j_z}c_{\mu s'_z}^{+}b_{\nu j_z}+h.c.$  and the Coulomb matrix element  $\langle f|V_{t~Coul}|i\rangle$  is equal to
$e^2/\epsilon_{\infty}\langle f|\hat {\bf d}\hat {\bf P}|i\rangle$, where  $\hat {\bf d}=\sum_{m, m'} {\bf d}_{ms_zm's_z}d_{ms_z}^+d_{m' s_z}$ is the operator of the intraionic dipole transition, and  $\hat {\bf P}=\sum_{\mu\nu s'_zj_z}{\bf P}_{\mu s'_z\nu j_z}c_{\mu s'_z}^+b_{\nu j_z}+h.c.$.

The matrix element
\begin{equation}
{\bf P}_{\mu s'_z\nu j_z}=\langle \mu s'_z|\frac{({\bf r}_2-{\bf a}_0)}{|{\bf r}_2-{\bf a}_0|^3}|\nu j_z\rangle=
\end{equation}
$$
-\frac{4\pi i}{(2\pi)^3} \int \frac{d^3 q~{\bf q}}{q^2}  \langle \mu s'_z|e^{i{\bf q(r_2-a_0)}}|\nu j_z\rangle
$$
The wave-functions of the ground state of conduction electron,$\mu=0$, are  $\langle {\bf r}|0 s'_z=\pm 1/2\rangle=\langle {\bf r}|c^+_{\pm 1/2}|0\rangle=\varphi^e_{\pm 1/2}({\bf r})=F_{es0}({\bf r})u_{c,\pm 1/2}({\bf r})=\sum_{\bf k} A_c({\bf k})e^{i{\bf kr}}u_{c,\pm 1/2}({\bf r})$, where $u_{c,\mp 1/2}({\bf r})=S_{\mp 1/2}$.

The weak admixture of valence band states to conduction ones is unimportant in calculation of ${\bf P}_{\mu s'_z\nu j_z}$ and can be omitted. The wave-functions of the ground state of valence electron, $\nu=0$, are
$\langle {\bf r}|0 j_z=\pm 3/2\rangle=\varphi^v_{\pm 3/2}({\bf r})=\langle {\bf r}|b^+_{\pm 3/2}|0\rangle=F^*_{hh0}({\bf r})u_{v,\pm 3/2}({\bf r})=\sum_{\bf k} A_v({\bf k})e^{i{\bf kr}}u_{v,\pm 3/2}({\bf r})$, where $u_{v,\pm 3/2}({\bf r})=\mp (X+iY)_{\pm 1/2}/\sqrt{2}$ are periodic parts of Bloch functions of conduction and valence  electrons, respectively. Wave-functions of the bright exciton are $|\pm 1\rangle=c_{\mp 1/2}^+b_{\pm 3/2}|0\rangle$. ${\bf P}_{0 s'_z 0 j_z}=-4\pi i F^*_{es0}({\bf a}_0)F^*_{hh0}({\bf a}_0){\bf R}_{s'_zj_z}$
because
\begin{equation}
P_{i~0 s'_z0 j_z}=\langle 0 s'_z|r_{2j}|0 j_z\rangle \frac{4\pi}{(2\pi)^3} \int \frac{d^3 q~ q_iq_j}{q^2}
\end{equation}
Here the relation $\langle n_in_j\rangle=4\pi/3\delta_{ij}$ for the angle integration is used, where ${\bf n}$ is the unit length vector. The vector
\begin{equation}
{\bf R}_{s'_zj_z}=\langle u_{c,s'_z}|{\bf r}|u_{v,j_z}\rangle=\frac{1}{\Omega_0}\int_{\Omega_0}d^3{r}
u^*_{c,s'_z}({\bf r}){\bf r}u_{v,j_z}({\bf r}),
\label{Beatti}
\end{equation}
where $\Omega_0$ is the unit cell volume, can be ${\bf R}_{\pm 1/2 \pm 3/2}=\mp 1/\sqrt{2}({\bf r}_{x~cv} \pm i{\bf r}_{y~cv})=\mp r_{cv}1/\sqrt{2}({\bf e}_x\pm i{\bf e}_y)$. Here ${\bf e}_x,{\bf e}_y$ are {\bf 0x} and {\bf 0y} unit vectors, ${\bf r}_{x~cv}=\langle S|{\bf r}|X\rangle$, ${\bf r}_{y~cv}=\langle S|{\bf r}|Y\rangle$, $r_{cv}=\langle S|y|Y\rangle=\langle S|x|X\rangle=\langle S|z|Z\rangle$.

The Coulomb matrix element between the initial
$|i(5/2,S_z,G)\rangle=^6\hat A_1(5/2,S_z)\hat \psi_{ex}(G)|0\rangle$ and the final  $|f(3/2,S'_z)\rangle=^4 \hat T_1(3/2,S_z)|0\rangle$ states is $M_{Coul \gamma S'_z,S_z}= -e^2/\epsilon_{\infty}{\bf d}_{r~\gamma S'_z,S_z} \langle 0|\hat {\bf P}|\psi_{ex}(G)\rangle$, where  ${\bf d}_{r~\gamma S'_z,S_z}=\langle T_{1\gamma}(S'_z)|\hat {\bf d}|A_1(S_z)\rangle$ is the matrix element of the $^6A_1(S_z)\to ^4T_{1\gamma}(S'_z)$ dipole transition. The transition is spin and parity forbidden so that the spin-orbit interaction and admixture of {\it p} states to $d_{ms_z}$ should be taken into account otherwise $d_{r\gamma S'_z,S_z}=0$. According to the model of Boulanger {\it et al.}  \cite{Boulanger} the optical  transition $^6A_1\to ^4T_{1\gamma}$ takes place by means of spin-orbit interaction via intermediate state $^4\tilde T_1(3/2)$ that allows for the {\it p-d} mixing.

Because of the symmetry $d_{rx~\gamma S'_zS_z}r_{x~cv}=d_{ry~\gamma S'_zS_z}r_{y~cv}=d_{rz~\gamma S'_zS_z}r_{z~cv}$ and $({\bf d}_{r~\gamma S'_zS_z}{\bf r}_{cv})^2=r_{cv}^2|{\bf d}_{r~\gamma S'_zS_z}|^2$. The matrix element of transition $^6\hat A_1(5/2,S_z)\hat \psi_{ex}(J_z=\pm 1)|0\rangle\to$ $^4 \hat T_{1\gamma}(3/2,S'_z)|0\rangle$ is
\begin{equation}
M_{Coul~\gamma S'_z S_z}=\pm \frac{4\pi e^2}{\epsilon_{\infty}}F_{es0}({\bf a}_0)F_{hh0}({\bf a}_0){\bf d}_{r~\gamma S'_z,S_z}{\bf r}_{cv}\frac{1\mp i}{2\sqrt{2}},
\end{equation}
which means that the Coulomb process reduces to the dipole-dipole energy transfer.

The above result can be reproduced if the matrix element $M_{Coul~\gamma S'_z,S_z}=\langle ^4T_{1\gamma}(3/2,S'_z) |V_{sc}| ^6A_1(5/2,S_z)\psi_{ex}\rangle$ instead of $\langle f|\hat V_{t~Coul}|i\rangle $ is considered, where
$V_{sc}({\bf r}, {\bf r}_1...{\bf r}_5 )=\sum_{i=1}^5 v_{sc}({\bf r}_i-{\bf r})$. Indeed, the matrix element of transition $|^6 A_1(5/2,S_z)\psi_{ex}(J_z=\pm 1)\rangle\to$ $|^4 T_{1\gamma}(3/2,S'_z)\rangle$ can be expressed as
$$
M_{Coul~\gamma S'_zS_z}=
\sum_{{\bf k},{\bf k}'}\langle A_c({\bf k})A^*_v({\bf k}')u^*_{v,j_z}({\bf r})
$$
$$
e^{i({\bf k}-{\bf k}'){\bf
r}} {^4T^*_{1\gamma}(3/2,S'_z)}|V_{sc}|u_{c,s_z}({\bf r})^6A_1(5/2,S_z)\rangle=
$$
\begin{equation}
\frac{4\pi e^2}{\epsilon_{\infty}}\frac{1}{(2\pi)^3}\sum_{{\bf k},{\bf k}'} A_c({\bf k})A^*_v({\bf k}')e^{i({\bf k}-{\bf k}'){\bf a}_0}
\end{equation}
$$
\times\int d^3{q}\frac{1}{{\bf q}^2}
J_{s_z{\bf k},j_z{\bf k}'}({\bf q})b_{\gamma S'_z,S_z}({\bf q}),
$$

where
$b_{\gamma S'_z,S_z}({\bf q})=\sum_i\int ^4T^*_{1\gamma}(3/2,S'_z)e^{-i{\bf q(r_i-a_0)}}$
${^6A_1}(5/2,S_z)d^3{ r}_i\simeq -i{\bf q}{\bf d}_{r~\gamma S'_zS_z}$, because
$\sum_i\int {^4T}^*_{1\gamma}(3/2,S'_z){\bf r}_i{^6A}_1(5/2,S_z)d^3{r}_i={\bf d}_{r~\gamma S'_zS_z}$.
The matrix element $ J_{s_z {\bf k},j_z{\bf k}'}({\bf q})=\langle u^*_{v,j_z}({\bf r})| e^{i({\bf k}-{\bf k}'+{\bf q}){\bf (r-a_0)}}|u_{c,s_z}({\bf r})\rangle= i({\bf k}-{\bf k}'+{\bf q})V{\bf R}^*_{s_zj_z}$.
With that, the matrix element $M_{Coul~\gamma S'_zS_z}$ becomes:
$$
M_{Coul~\gamma S'_zS_z}=\pm \frac{4\pi e^2}{\epsilon_{\infty}}\sum_{{\bf k},{\bf k}'}{\bf d}_{r~\gamma S_zS'_z}{\bf r}_{cv}A_c({\bf k})
$$
$$
\times e^{i{\bf ka_0}}e^{-i{\bf k'a_0}}A^*_v({\bf k}')\frac{1\mp i}{2\sqrt{2}}=
$$
\begin{equation}
\pm \frac{4\pi e^2}{\epsilon_{\infty}}F_{es0}({\bf a}_0)F_{hh0}({\bf a}_0) {\bf d}_{r~\gamma S'_zS_z} {\bf r}_{cv}\frac{1\mp i}{2\sqrt{2}},
\label{Coulomic}
\end{equation}

Matrix elements ${\bf d}_{r~\gamma S'_zS_z}$ are related to the radiative life-time of $^4T_1\to ^6A_1$ transition:
\begin{equation}
\frac{1}{3}\sum_{\gamma}\sum_{S'_zS_z}|{\bf d}_{r~\gamma S'_zS_z}|^2=\frac{3}{4}\frac{N_i}{n_r}\frac{\hbar^4 c^3}{e^2E^3_0}\frac{1}{\tau_r}=\frac{3}{2}D_r^2,
\end{equation}
where $\tau_r$ is the radiative time, $N_i=(2\times 5/2+1)$ is the degeneracy of the state $^6A_1(5/2,S_z)$, $n_r=\sqrt{\epsilon_0}\simeq 3$ is the optical refraction index, and $E_0\simeq 2.15~eV$ is the transition energy; $(D_r/a_B)^2\simeq 10^{-4}$ for CdMnSe  since $\tau_r=200$ $\mu$s. \cite{Agekyan} The factor 3/2 appeares because only $d_{rx~\gamma S'_zS_z}$ and $d_{ry~\gamma S'_zS_z}$ components are involved in the nonradiative recombination; the factor 1/3 is because of degeneracy of the $T_{1\gamma}(3/2,S_z)$ state.

Summation over Mn$^{2+}$ ions interacting with the bright exciton gives the rate of the transition $(|^6A_1\psi_{ex}\rangle\to |^4T_1\rangle)$:
\begin{equation}
R_{Coul}=\frac{2\pi}{\hbar}(4\pi)^2\frac{1}{N_i}\left
(\frac{e^2}{\epsilon_{\infty}
a_B}\right)^2a^4_BD_r^2 \frac{\hbar^2}{8m_ca_B^2E_g}
\end{equation}
$$
\times\sum_\lambda|F_{es0}({\bf a}_\lambda)|^2|F_{hh0}({\bf a}_\lambda)|^2\delta(E_i-E_f),
$$

where $a_B$ is the atomic Bohr radius, and $\lambda$ enumerates Mn$^{2+}$ ions. In derivation of this equation relations $r=i\hbar p/m_0E_g$, $p=iPm_0/\hbar$,  and $P^2=3E_g(\Delta+E_g)\hbar/(2(2\Delta+3E_g)m_c)\simeq E_g\hbar^2/2m_c$ are used, where $m_c=0.13m_0$ is the conduction electron mass in CdSe or ZnSe; \cite{Abakumov} $P$ is the Kane constant. Since $\Delta/E_g<0.3$ in ZnSe and CdSe crystals we assume that $\Delta/E_g=0$.

After averaging over positions of Mn$^{2+}$ ion, the rate becomes:
\begin{equation}
\overline{R}_{Coul}=\frac{2\pi}{\hbar}(4\pi)^2N_{Mn}\frac{1}{6}\left (\frac{e^2}{\epsilon_{\infty}
a_B}\right)^2\frac{a^6_B}{V_eV_h}\eta_e\eta_h
\end{equation}
$$
\times\left(\frac{D_r}{a_B}\right)^2
\frac{\hbar^2}{8m_ca^2_BE_g}\delta(E_i-E_f),
$$
where $V_{e(h)}$ is the effective volume of electron (hole) localization and coefficients $\eta_{e(h)}=\int_{DMS} d^3 r|F_{e(h)}(r)|^2$  characterize penetration of the electron(hole) wave-function into the DMS layer.

The presented analysis  generalizes the approach developed in Ref. \cite{Robbins,Allen86,Yassievich93}, where the interconfigurational optical transitions are considered as single-electron ones.

\section{Matrix element of nonradiative recombination due to sp-d hybridization}

The Hamiltonian responsible for the nonradiative recombination via {\it sp-d} mixing is
\begin{equation}
\hat H_{mix}=\sum_{m, m',\nu,\mu,s_z,s'_z,j_z} (K_{mix}d^+_{ms_z}d_{m's'_z}b^+_{\nu j_z}c_{\mu s_z} +h.c.),
\end{equation}
where
\begin{equation}
K_{mix}=-V_{sd~ ms_z\mu s_z}V^*_{pd~ m's'_z\nu j_z}\left(\frac{1}{E_i-E^+}+\frac{1}{E_i-E^-}\right)
\end{equation}
and $E_i$ is the energy of the initial state whereas $E^+$ and $E^-$ are energies of intermediate states. For calculations of the matrix element we use wave-functions of the ground and excited states in the spherical approximation similar to the approach of Schriffer in the calculation of the kinetic exchange constant. \cite{Schriffer67} We choose the initial state of the Mn$^{+2}$ ion to be $|A_1(5/2,-5/2)=\Pi_{m=-2}^2 d^+_{m~-1/2}|0\rangle$. States with other spin projections can be obtained from this state by means of the step-up $\hat S_+=\sum_m d^+_{m~1/2}d_{m~-1/2}$ and step-down $\hat S_-=\sum_m d^+_{m~-1/2}d_{m~1/2}$ spin operators: $|5/2,S_z\rangle=C\hat {S}^{5/2-S_z}_{-}\Pi_{m=-2}^2 d^+_{m~1/2}|0\rangle$, where $C$ is the normalization constant. Particularly, $C=1/\sqrt{5}$ for $S_z=\pm 3/2$, and  $C=1/\sqrt{10}$ for $S_z=\pm 1/2$.

We assume here that Hund's rules are valid. Thus, the final state is $G^4(3/2)$ excited state of Mn$^{2+}$ ion $| 3/2, -3/2\rangle=d^+_{2~1/2}d^+_{2~-1/2}\Pi_{m=-1}^1 d^+_{m~1/2}|0\rangle$. Creation and annihilation operators $d^+_m$, $d_m$ satisfy standard commutational relations. Although $\langle \varphi^e|d_m\rangle\neq 0$ and  $\langle \varphi^v|d_m\rangle \neq 0$, Kane basis functions are orthogonal to {\it 3d}-states, so that the nonorthogonality is negligibly small.

The initial state for the recombination of bright exciton $|1\rangle=c^+_{-1/2} b_{3/2}|0\rangle$ is
$|i\rangle=c^+_{-1/2} b_{3/2}C\hat {S}^{5/2-S_z}_{-}\Pi_{m=-2}^2 d^+_{m~1/2}|0\rangle$,
whereas the final state is  $|f\rangle=d^+_{2~1/2}d^+_{2~-1/2}C'\hat S^{3/2-S_z}\Pi_{m=-1}^1 d^+_{m~1/2}|0\rangle$.

Since $V_{pd}$ couples valence band states with {\it 3d}-states of $t_2$ representation it is useful to express the former via states of tetrahedral symmetry: $d_{\pm 2}=1/\sqrt{2}(d_{ev}-d_{t_2~\zeta})\propto(x\pm iy)^2$, $d_{\pm 1}=\mp 1/\sqrt{2}(d_{t_2\zeta}\pm id_{t_2\xi})\propto y(x\pm iz)$, $d_0=d_{eu}\propto (3z^2-r^2)$. Here $\xi, \zeta,\eta$ are basis functions of the $t_2$ representation, $u$ and $v$ are basis functions of $e$ representation of the tetrahedral group. \cite{PRB05,Griffith}  Matrix elements of the {\it s-d} coupling are  $|V_{sd~\pm 2}|=|V_{sd~\pm 1}|=|V_{pd}|\sqrt{\gamma/2}$. Transitions involving the state $d_0$ are forbidden because it does not mix with valence band states, i.e. $V_{sd~0}=0$.

By means of the relation $V_{pd}=4V^0_{pd}/\sqrt{N_0}$, where $V^0_{pd}$ is the hopping amplitude \cite{Larson88,Bhatt92} the matrix element of the transition $|A_1(\pm 3/2)J_z=1\rangle\to |T_1(\pm 3/2)\rangle$ ($|A_1(\mp 3/2)J_z=- 1\rangle\to |T_1(\mp 3/2)\rangle$) is
\begin{equation}
M_{mix}= \alpha(\pm 3/2) \frac{16}{N_0}F_{ez0}({\bf a}_0)F_{hh0}({\bf a}_0)|V^0_{pd}|^2
\label{Mif}
\end{equation}
$$
\times\left(\frac{1}{U_{eff}+\epsilon_d-E_c}+\frac{1}{E_v-\epsilon_d}\right)\left(\frac{\gamma}{2}\right)^{1/2},
$$
where $N_0$ is the number of unit cells per unit volume and $\alpha(\pm 3/2)=-1/\sqrt{5}$. By means of spin-up  operators and wave-functions of initial and final states the coefficient $\alpha(\pm 1/2)=-\sqrt{3/10}$ for the transition $|A_1(\pm 1/2)J_z=1\rangle\to |T_1(\pm 1/2)\rangle$ can be found.

By using $\varphi_{s_z}^e(k)$ for free electrons the coefficient $\gamma$ is evaluated as $\gamma\simeq \hbar^2/2m_cE_gL^2$. Smaller value $\gamma\simeq \hbar^2/2m_cE_gL^2/(1+\hbar^2/2m_cE_gL^2)$ can be found for the symmetric quantum well of width $L$ with infinitely high barriers by using functions $\varphi^e_{\pm 1/2}({\bf r})$ presented in Ref.\cite{Merkulov00}. Since $E_i=5\epsilon_d-E_v+E_c$, $E^+=6\epsilon_d+U_{eff}-E_v$, $E^-=4\epsilon_d+E_c$, $U_{eff}=7.0$ eV then $E_i-E^-=-(E_v-\epsilon_d)\simeq -3.5$ eV, $E_i-E^+=-(U_{eff}+\epsilon_{d}-E_c)\simeq -1.5$ eV. Values of  $U_{eff}$ , $E_v-\epsilon_d$  and $V^0_{pd}\simeq -0.6$ eV are taken from Ref.\cite{Young91}

Taking into account Eq.(\ref{Mif}) the probability of nonradiative recombination is as follows:
\begin{equation}
R_{mix}=\frac{1}{\tau_{mix}}=\frac{1}{N_i}\sum_{\lambda}\sum_{if}\frac{2\pi}{\hbar}|M_{mix~\lambda}|^2\delta(E_i-E_f)
\end{equation}
Here $\lambda$ enumerates Mn$^{2+}$ ions. The recombination rate of the bright exciton averaged over the Mn$^{2+}$ ions distribution can be obtained in the manner described in Appendix A.

The ratio of rates, corresponding to the Coulomb and {\it sp-d} mixing processes is:
\begin{equation}
\frac{\overline {R}_{Coul}}{\overline{R}_{mix}}= \frac{1}{2}\left(\frac{e^2}{V^0_{pd}\epsilon_{\infty}a_B} \right)^2 \left(\frac{a_B^3}{\Omega_0}\right)^2
\left(\frac{D_r}{a_B}\right)^2
\end{equation}
$$
\left(\frac{1}{UV_{pd}}\right)^2\left(\frac{L}{a_B}\right)^2\simeq 10^{-2},
$$
where $U=1/(\epsilon_d+U_{eff}-E_c)+1/(E_v-\epsilon_d)$, $\epsilon_{\infty}\simeq 6$, $\Omega_0$ is the volume of unit cell.

The parameters of $U_{eff}$, $E_v-\epsilon_d$,  and $V^0_{pd}$ are not independent but related to each other via the exchange constant $\beta$.  These parameters are obtained by means of theoretical and numerical analysis of Zeeman splitting of band states and photoemission spectroscopy data. \cite{Larson88} The experimental results of Zeeman splitting as well as photoemission data are reported with accuracy of two significant figures. The best correspondence between experimental and theoretical results are found with values $U_{eff}=7.0$ eV, $E_v-\epsilon_d=3.5$ eV, $V^0_{pd}=0.6$ eV for CdMnSe. \cite{Larson88,Young91} Taking into account uncertainty of the parameter $\epsilon_d+U_{eff}-E_c$ related to the fact that $E_g$ depends on the Mn content and strain distribution within the sample we conclude that the accuracy of our estimate is better than an order of magnitude.

\newpage

\begin{figure}
\begin{center}
\includegraphics[width=8.5cm]{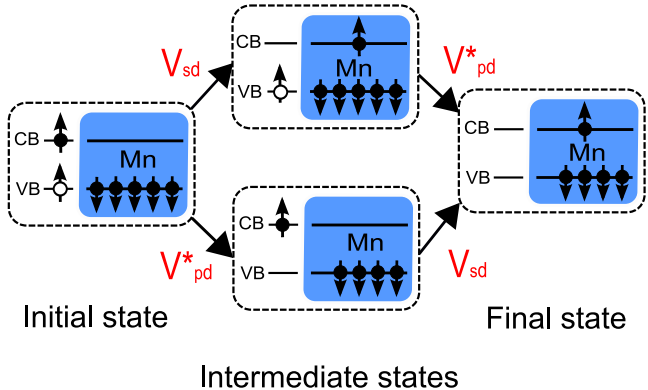}
\end{center}
\caption{The scheme illustrates the process of the nonradiative recombination of the dark exciton $J_z=2$ in a magnetic field ${\bf B}\parallel {\bf 0z}$. Two different paths of the process via virtual states contributing to (\ref{Mif0}) are shown. The {\it sp-d} mechanisms of exciton recombination can be described as the result of successive hopping of the electron(hole) and hole(electron) in the {\it 3d}-shell.} \label{T}
\label{Ex}
\end{figure}

\begin{figure}
\begin{center}
\includegraphics[width=8.5cm]{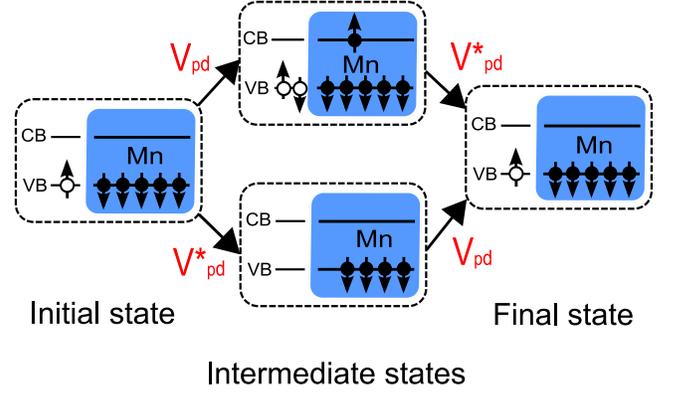}
\caption{The scheme illustrates processes of the kinetic {\it p-d} exchange interaction between {\it hh}  $|3/2\rangle$ state and Mn$^{2+}$ {\it 3d}-shell. Two different paths of transitions via virtual intermediate "donor" $d^4$ and "acceptor" $d^6$  states are shown.}
\end{center}
\end{figure}

\end{document}